\documentclass[ twocolumn,
superscriptaddress,
 amsmath,amssymb,
 aps, prd,
]{revtex4-2}
\usepackage{graphicx} 
\usepackage{hyperref}
\usepackage{orcidlink}
\usepackage[justification=justified,singlelinecheck=false]{caption}
\usepackage{subcaption}
\usepackage{amsmath}
\usepackage[english]{babel}

\usepackage{xcolor}

\newcommand{\AEI}{\affiliation{Max Planck Institute for Gravitational Physics (Albert Einstein Institute), Am M\"uhlenberg 1, Potsdam 14476, Germany}}

\begin{document}

\title{Dissecting and diagnosing the LISA Galactic binary population}
\author{Alice E. Cravioglio}
\email{alice.cravioglio@aei.mpg.de}
\AEI
\affiliation{Institute for Particle Physics and Astrophysics, ETH Zurich, Zurich, Switzerland}

\author{Arianna I. Renzini \orcidlink{0000-0002-4589-3987}}
\email{arenzini@ethz.phys.ch}
\affiliation{Institute for Particle Physics and Astrophysics, ETH Zurich, Zurich, Switzerland}

\begin{abstract}
LISA will individually resolve thousands of Galactic double white dwarf (DWD) binaries, while the majority stays buried in an unresolved gravitational wave (GW) confusion foreground. Which systems get resolved depends on the binary's parameters, on the global population and on instrumental sensitivity, so the resolved population is not a representative sample of the underlying Galactic population. Interpreting LISA's future catalog therefore requires understanding these selection effects. Here we characterize the resolved and unresolved subpopulations of a simulated Milky Way DWD catalog, using an iterative subtraction pipeline applied over different observation times. Resolvability is driven primarily by chirp mass: higher chirp mass, inspiraling systems are strongly favored over lower mass, outspiraling or monochromatic ones, and this reflects on the rate with which different composition subtypes are resolved. Spatially, resolved sources trace the Galactic bulge, because it is intrinsically the densest region of the Galaxy. The confusion background shows an internal bimodal chirp-mass structure, and is not generated uniformly by the same subpopulation in different frequency bands. We introduce a diagnostic for the minimum number of sources needed to reproduce a given fraction of the confusion power in a frequency band, and show that a residual confusion floor persists even after 100 years of integration. These biases are significant and need to be accounted for when using LISA's resolved DWD catalog and the confusion background to constrain the Galactic DWD population.
\end{abstract}

\maketitle

\section{Introduction}
\

The Laser Interferometer Space Antenna (LISA)~\cite{amaro-seoaneLaserInterferometerSpace2017} mission will open the mHz gravitational-wave (GW) band, providing access to a rich variety of astrophysical sources currently inaccessible to ground-based detectors \cite{colpiLISADefinitionStudy2024}. Among these, double white dwarfs (DWDs) in the Milky Way represent the most numerous source class: population synthesis models predict $\mathcal{O}(10^7)$ sources emitting in the LISA band at the same time \cite{lambertsPredictingLISAWhite2019, nelemans_galactic_2009, toonen_supernova_2012}.  The majority will be individually unresolvable, and they will overlap in time and frequency, forming a Galactic confusion foreground that constitutes both a noise source and a probe of the Galactic binary population \cite{nelemans_galactic_2009, hils_gravitational_1990, timpanoCharacterizingGalacticGravitational2006}. A fraction of these binaries are expected to be detectable and characterizable using parameter estimation techniques \cite{crowder_solution_2007, nelemans_galactic_2009, korol_prospects_2017, littenberg_detection_2011}.

This operational division into two source categories, i.e. resolvable and unresolvable, raises a fundamental question for LISA data analysis and astrophysical inference: whether the resolved sources and the confusion foreground constitute two ends of the same underlying binary distribution, or whether they have different statistical properties. If the resolved and unresolved systems are drawn from the same underlying distribution, the first ones could constrain the properties of the population contributing to the foreground; if they differ systematically, extrapolating from the resolved catalog to the unresolved set could bias population inference. A related but distinct question is whether the confusion foreground is itself produced representatively by the unresolved population, or whether a comparatively small subset of systems dominates its contribution to the total background power. In other words, being unresolvable and contributing significantly to the confusion noise level are not necessarily the same property, and it is this distinction we aim to characterize.

Assessing this requires a systematic characterization of the salient features of each population independently. Specifically, we ask three related questions. First, what physical and spatial properties characterize the systems that dominate the confusion background? Second, how do these systems compare with individually resolved DWDs? Third, to what extent does the resolved population reflect the true underlying Galactic DWD population? Investigating these aspects will inform expectations for DWD science with LISA and provide guidelines for parameter-inference strategies.

In this work we address these questions through an extensive analysis of a newly simulated catalog of Galactic DWDs, generated from the same cosmological simulation framework used in the LISA common dataset challenges and therefore representative of the current gold standard for LISA population modeling \cite{baghi_lisa_2022, babak_mock_2010}. Applying the iterative subtraction logic as in Karnesis et al. \cite{karnesisCharacterizationStochasticSignal2021}, for observation times of 1, 2 and 4 years, we systematically characterize both the resolved population and the confusion foreground across their distribution in parameter space, their compositional fractions, their spatial distribution, and the extent to which the confusion itself is concentrated among a subset of the unresolved sources. The selection effects we uncover, such as the preference for higher chirp mass and bulge's dominance among resolved sources, follow from general physical arguments and should be robust. The specific compositional trends, however, are a feature of this particular simulated population and depend on its binary-evolution assumptions, so they should not be read as ground truth for the real Milky Way.

Beyond this time-averaged characterization, the confusion foreground is not stationary in time: due to the orbit of LISA around the Sun and to the anisotropic distribution of the Galactic binaries, a time-dependent amplitude modulation is induced on the observed signal \cite{seto_annual_2004, digman_lisa_2022}. This cyclostationarity of the foreground carries information about the Milky Way structure, and introduces a way to distinguish it from the instrumental noise or other isotropic backgrounds, such as extragalactic DWDs \cite{pozzoli_cyclostationary_2025, farmer_gravitational_2003, staelensLikelihoodWhiteDwarf2024, adams_detecting_2014}. For this reason, a simplified analysis on the time-dependence of the foreground is also carried out. 

This paper is organized as follows. Section \ref{sec:methods} describes the simulated catalog, the iterative subtraction pipeline, the minimum-source-count diagnostic used to quantify how concentrated the confusion power is and our treatment of the foreground's time-dependence. Section \ref{sec:results} presents our results on detectability and on the resulting background, discussing compositional and spatial biases in the resolved population; lastly, we discuss the time evolution of the confusion. Section \ref{sec:discussion} discusses the physical origin and implications of these biases, together with the limitations of this analysis, and summarizes our conclusions.

\section{Methods} \label{sec:methods}

\subsection{The source catalog}

The catalog used for the analysis is a realistic simulation of the ultra compact binaries in the Milky Way produced by A. Lamberts and A. Toubiana \cite{lambertsPredictingLISAWhite2019, toubianaInteractingDoubleWhite2024}. It is a different realization of the same underlying simulation used in Lamberts et al. \cite{lambertsPredictingLISAWhite2019}, generated with different initial conditions \cite{wetzel_reconciling_2016, hopkins_fire-2_2018}. It contains $\sim$15 million sources, each characterized by the following set of binary parameters: frequency $f$, frequency derivative $\dot{f}$, strain amplitude $A$, sky location ($\alpha, \delta$), inclination $\iota$, polarization angle $\psi$ and initial phase $\phi$. In addition to the standard Galactic binary parameters, the catalog contains their luminosity distance $d_L$ and the core composition of the WDs in the system: He (Helium), CO (Carbon/Oxygen) or Ne (Neon) WDs. These types originate from different progenitor masses and/or binary evolution channels, and contribute differently to the GW population \cite{lambertsPredictingLISAWhite2019}. Depending on the nature of its WDs components, there are 4 types of binaries: He-CO, He-He, CO-CO and Ne-X, where X can be any type of WD. He-He binaries are characterized by low chirp masses, tighter orbits and a slow evolution, CO-CO systems come from initially wider orbits and have masses above $0.45 \ M_\odot$; He-CO WDs have lower chirp masses due to the He component and are the most numerous in the catalog. Finally, Ne-X WDs stem from massive stars, they form on short timescales and are rare \cite{lambertsPredictingLISAWhite2019, toonen_supernova_2012, nelemans_galactic_2009}. Binary type is therefore an approximate proxy for chirp mass. These four compositional subtypes follow directly from WD formation channels and are expected in any population-synthesis model; the relative fraction of each subtype, however, is sensitive to the specific binary-evolution assumptions adopted.
The chirp mass is also one of the primary drivers of detectability, since the GW strain amplitude of a quasi-monochromatic DWD scales as:
\begin{equation}
    A \propto \frac{\mathcal{M}_c^{5/3} f^{2/3}}{d_L} ,
    \label{eq:amplitude-chirp-mass}
\end{equation}
so that, at fixed frequency and distance, systems with larger chirp masses produce intrinsically stronger signals.

The catalog's host galaxy is drawn from a cosmological simulation of a Milky-Way-mass galaxy from the FIRE-2 project \cite{hopkins_fire-2_2018}, which is why it naturally includes realistic substructure, such as stellar streams and satellite galaxies, alongside the smooth disk and bulge, and extends up to 300 kpc from the Sun. We follow Lamberts et al. \cite{lambertsPredictingLISAWhite2019} in defining the halo sources as sources within the virial radius of the main galaxy that did not form in it: this includes satellite galaxies, stellar streams and remnants of disrupted satellites, but excludes stars that were formed in the main galaxy that were perturbed onto more radial orbits. About 5\% of the binaries in the catalog are part of the halo. 

The main difference with respect to Lamberts et al.'s catalog \cite{lambertsPredictingLISAWhite2019} is the inclusion of mass transferring binaries, which are characterized by a negative frequency derivative ($\dot{f}<0$), meaning they are outspiraling. Toubiana et al. \cite{toubianaInteractingDoubleWhite2024} took the same simulations of Lamberts et al. \cite{lambertsPredictingLISAWhite2019} and accounted for the effect of mass transfer and tides on the evolution of the binaries, leading to outspiraling binaries.

\begin{figure}[h!] 
    \centering
    \begin{subfigure}[b]{0.49 \textwidth}
        \includegraphics[width =\textwidth]{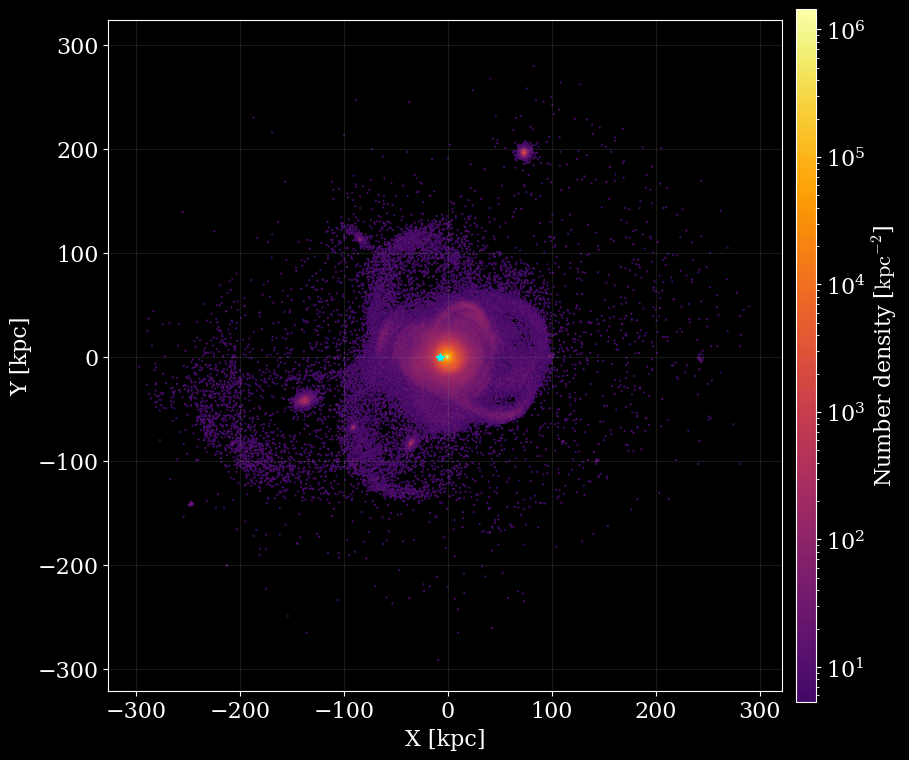}
    \end{subfigure}
    \begin{subfigure}[b]{0.49 \textwidth}
        \includegraphics[width=\textwidth]{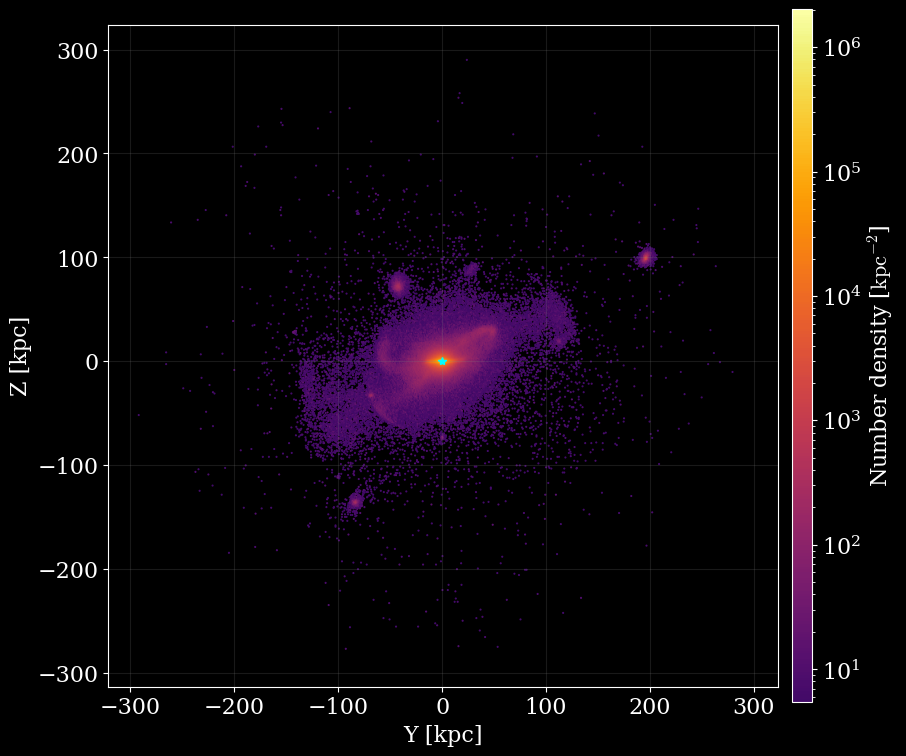}
    \end{subfigure}
    \caption{View of the number density of the sources face-on (XY-plane, top panel) and edge-on (YZ-plane, bottom panel). The cyan star indicates the position of the Sun.}
    \label{fig:density-catalog-comp-all-XY-YZ}
\end{figure}

\begin{figure*}[ht]
    \centering
    \includegraphics[width=0.8\linewidth]{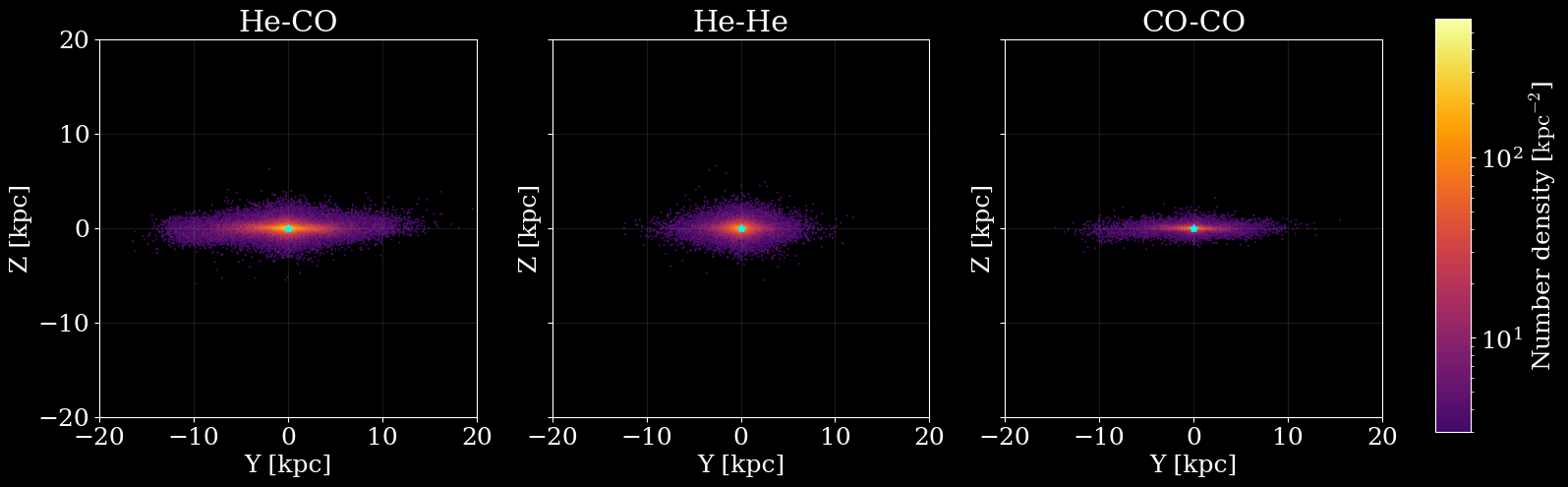}
    \caption{Density maps of the He-CO DWDs (left), He-He DWDs (middle) and CO-CO DWDs (right) from a side view. The binaries shown are only the ones with $f\geq 1$ mHz, which are the most likely resolvable from LISA. The Ne-X binaries are not shown due to their rarity, but their distribution is similar to the He-He DWDs.}
    \label{fig:density-types-zoom}
\end{figure*}

Figure \ref{fig:density-catalog-comp-all-XY-YZ} shows the surface density of the binaries in the catalog, with a face-on and edge-on view. The dense Galactic bulge and the disk are well visible, as well as stellar streams and some satellite galaxies. Figure \ref{fig:density-types-zoom} shows the spatial distribution viewed edge-on of the different binary types, for the highest frequency systems ($f \geq 1$ mHz), which are more likely to be detected by LISA. These are concentrated in the bulge and disk, and different subtypes have slightly different distributions, with CO-CO binaries present almost exclusively in the disk and He-He DWDs spread around the bulge and halo in a more spherical distribution. The He-CO binaries follow an intermediate distribution, with a more prominent disk and some bulge contribution.

\subsection{Iterative subtraction of loud sources}

To estimate the confusion noise from the unresolved binaries and extract the resolvable sources, an iterative subtraction procedure was implemented following the approach described by Karnesis et al. \cite{karnesisCharacterizationStochasticSignal2021}. The method aims to iteratively separate individually resolvable sources from the unresolved population contributing to the confusion, based on a SNR-calculation. The instrumental noise models for the acceleration noise $S_\mathrm{acc}$ and the optical metrology system (OMS) noise $S_\mathrm{OMS}$ used by the pipeline are the following \cite{robsonConstructionUseLISA2019}:

\begin{equation}
    S_{\mathrm{OMS}} = (1.5 \times 10^{-11} \: \mathrm{m })^2 \left( 1 + \left( \frac{2 \: \mathrm{mHz}}{f} \right)^4 \right) \mathrm{Hz}^{-1}
\end{equation}

\begin{align*}
    S_{\mathrm{acc}} = \left(3 \times 10^{-15} \: \frac{\mathrm{m}}{\mathrm{s}^2}\right)^2 \left( 1 + \left( \frac{0.4 \: \mathrm{mHz}}{f} \right)^2  \right) \times \\ \left(1 + \left( \frac{f}{8 \: \mathrm{mHz}} \right)^4 \right)\mathrm{Hz}^{-1}
\end{align*}
These are combined to obtained the total instrumental noise PSD in the TDI A and E channel as follows \cite{prince_lisa_2002, tinto_time_2004}:
\begin{align*}
    S_{\mathrm{instr}} = 16 \Lambda(f) \sin^2(\omega \mathrm{L}) \left[ 3 + 2 \cos(\omega \mathrm{L}) + \cos(2 \omega \mathrm{L}) \right] S_{\mathrm{acc}}  +\\
    +  8 \Lambda(f) \sin^2(\omega \mathrm{L}) \left[  2 + \cos(\omega \mathrm{L}) \right] S_{\mathrm{OMS}}
\end{align*}
where $\omega = 2\pi f$, L is the arm length of LISA in seconds and $\Lambda(f)$ depends on the TDI generation (1.5 or 2.0): $\Lambda_{1.5}(f) = 1$ and $\Lambda_{2.0}(f) = 4 \sin^2(2 \omega \mathrm{L})$.

For each binary, a waveform is generated over the full observation baseline $T_\mathrm{obs}$, including the LISA TDI response, yielding its frequency-domain strain $\tilde{h}(f)$ in the A and E channels. This is used directly in the optimal SNR integral below, without reducing to an analytic monochromatic approximation. The optimal SNR of a source emitting a signal $\tilde{h}(f)$ is then defined as:
\begin{equation} \label{eq:optimal_SNR}
    \mathrm{SNR}_{\mathrm{opt}}^2 = \left( \tilde{h}(f)|\tilde{h}(f)\right) = 4 \int_0^\infty df \, \frac{|\tilde{h}(f)|^2}{S_n^\mathrm{tot}(f)}
\end{equation}
where $S_n^\mathrm{tot}(f) = S_\mathrm{instr}(f) + S_\mathrm{conf}(f)$ is the sum of the instrumental noise and the confusion noise PSD from the currently unresolved sources.

At each iteration, $S_\mathrm{conf}(f)$ is recomputed from the sources not yet flagged as resolved; a source is classified as resolved once its SNR against the current $S_{n}^\mathrm{tot}(f)$ exceeds the chosen threshold, after which it is removed from the unresolved set and the procedure repeats until convergence. Since $\mathrm{SNR}\propto \sqrt{T_\mathrm{obs}} $, longer observation times directly improve detectability by accumulating signal. The package implemented for the subtraction is \verb|GBconfusion| \footnote{\url{https://github.com/dpcatch/GBconfusion}}: it takes as input the source catalog, and outputs a set of resolved sources, together with the residual noise PSD $S_n^\mathrm{tot}$. This iterative subtraction method is commonly used to estimate the Galactic confusion noise, as it yields results quickly and without requiring more complex analyses \cite{timpanoCharacterizingGalacticGravitational2006, karnesisCharacterizationStochasticSignal2021, cornish_galactic_2017}. However, it uses some simplifications, such as perfect knowledge of the data and of the noise, and perfect source subtraction. Given these limitations, this procedure should be taken as an optimistic upper limit to more accurate estimates.

The analysis was performed for the A and E TDI 2.0 channels, for observation times $T_\mathrm{obs} = [1, 2, 4]$ years, with the SNR threshold for detectability SNR=7.
The following analytical model was fitted to the output PSD \cite{karnesisCharacterizationStochasticSignal2021}, and the result was used to compute the final SNR of the sources:

\begin{equation}
    S_{\mathrm{gal}} = A  f^{-7/3} \exp  \left[ -(f/f_1)^\alpha \right] \left( 1 + \tanh\left( -\frac{f - f_k}{f_2}\right) \right)
\end{equation}
with
\begin{equation}
    \log(f_1) = a_1 \log(T_{\mathrm{obs}}) + b_1
\end{equation} 
\begin{equation}
    \log(f_k) = a_k \log(T_{\mathrm{obs}}) + b_k
\end{equation} 

\subsection{Minimum source count contributing to the confusion background} \label{sec:minimum-source-count-N99-A99}

The motivation for this diagnostic is best understood by analogy with generic stochastic background, for which the energy density is naturally written as an integral over source redshift: $\Omega_\mathrm{background} \propto \int_0^\infty dz (\mathrm{population})$. In that context, one can ask up to which redshift shell $z_x$ a fraction $x$ of the background accumulates, and compare it to the maximum redshift $z_\mathrm{max}$ at which individual sources are resolvable: if $z_\mathrm{max}<z_x$, the background signal is informative, and measuring it provides complementary information about the population. Redshift is not a meaningful ordering variable for Galactic DWDs, since these sources are local to the Milky Way. Amplitude plays the analogous role within each frequency band: each bandhas a minimum detectable amplitude $A_\mathrm{min}$ set by the resolvability threshold, and the unresolved sources contributing to the confusion can be ordered by decreasing amplitude to ask up to which threshold $A_x$ a fraction $x$ of the confusion power accumulates.

For each frequency band, the unresolved sources can be ranked by their individual contribution to the confusion power rather than simply counted. Given the $N$ unresolved sources with amplitudes $\{A_i\}$ in a given band, we order them by decreasing amplitude, $A_1 \geq A_2 \geq \dots \geq A_N$, and define the cumulative power

\begin{equation}
    S(k) = \sum_{i=1}^{k} A_i^2 , \qquad
    S_\mathrm{tot} \equiv S(N) = \sum_{i=1}^N A_i^2 ,
    \label{eq:cumulative-power}
\end{equation}

For a threshold $x \in \{50, 90, 95, 99\}$ (\%), we define $N_x$ as the smallest $k$ such that
\begin{equation}
    S(k) \geq S_\mathrm{tot}\,\frac{x}{100} ,
    \label{eq:threshold-definition}
\end{equation}

with corresponding amplitude threshold $A_x$: sources with $A \geq A_x$ account for $x \%$ of the confusion power in that band. $N_x$ is therefore the minimum number of sources required to reproduce a fraction $x$ of the confusion background in a given frequency band, effectively the smallest population capable of generating that noise level, since many different combinations of source amplitudes could otherwise yield the same noise. We apply this procedure with different thresholds throughout Section \ref{sec:results}, to assess which types of binaries dominate the background.

\subsection{Time-dependence of the foreground}

To quantify the yearly modulation of the foreground due to LISA's orbit, we start from the fixed set of unresolved sources obtained from the iterative subtraction procedure. We do not re-run the resolvability search at each epoch, since the goal is to characterize how the response to a given unresolved population varies in time.

For this set of sources, we generated the time-domain strain data for $T_\mathrm{obs} = 1$ year using \texttt{JaxGBaccurate} \footnote{\url{https://gitlab.com/lisamission/jaxgb}} \cite{cornishTestsBayesianModel2007}. Then, we divided the data into 52 one-week-long time chunks, and computed the confusion noise PSD for each week. Time chunks of one week are an optimal choice: it allows a frequency resolution $\Delta f = 1/T_\mathrm{chunk} \approx 1.65 \times 10^{-6}$ Hz, well below the typical frequencies of the binaries in the catalog, while remaining short enough that the background can be approximated as stationary within a single chunk.

We compare the resulting modulation to the one obtained from \verb|backgrounds| \footnote{\url{https://qbaghi.pages.in2p3.fr/backgrounds/index.html}}, which computes the LISA response analytically from only the amplitude and sky position of each source, without generating individual waveforms. This is effectively an averaged, fast estimate of the expected modulation shape. The comparison is therefore not a validation of one method against the other, but a check that the detailed, per-source recomputation with \verb|JaxGBaccurate| reproduces the overall modulation pattern already expected from the simpler analytic response.

\section{Results}
\label{sec:results}

In this section we review the following results: the properties of resolved sources (Section \ref{sec:detectability-bias}), the structure and origin of the confusion background (Section \ref{sec:confusion-background}), the spatial (Section \ref{sec:spatial-bias}) and compositional (Section \ref{sec:composition-bias}) biases of the resolved catalog with respect to the unresolved populations, and the time evolution over the year of the confusion (Section \ref{sec:time-evolution}).

\subsection{Properties of resolved sources} \label{sec:detectability-bias}

In this section we examine how chirp mass and frequency derivative jointly determine which DWDs are resolved, and how this detectability bias evolves with observation time.

As expected from Eq. \ref{eq:optimal_SNR}, $\mathrm{SNR}\propto\sqrt{T_\mathrm{obs}}$, and the number of resolved binaries increases with $T_\mathrm{obs}$ (see Table \ref{tab:mchirp-fdot-res-all}), from 4481 (1 yr) to 11811 (4 yr), corresponding to a factor of 2.6 and reflecting the improved sensitivity of the detector as signals accumulate over longer periods of time.

\begin{figure*}[ht]
    \centering
    \includegraphics[width=0.6\linewidth]{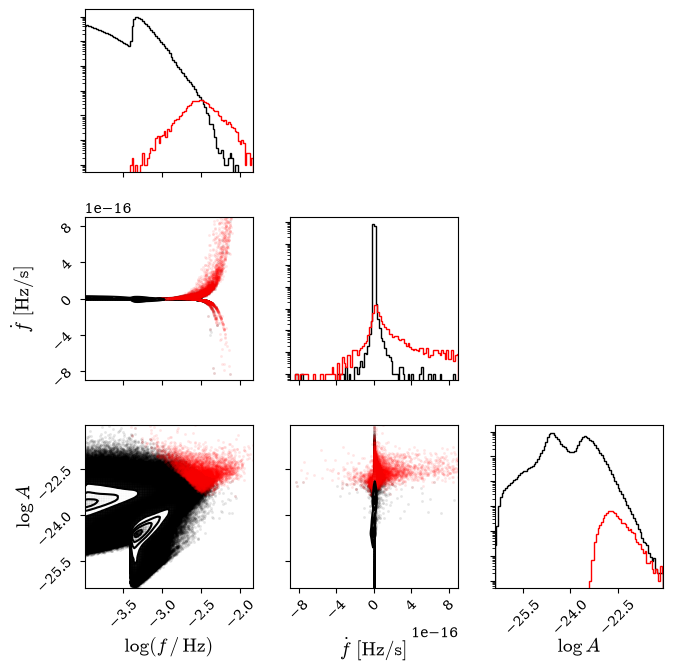}
    \caption{Parameter space plots of $f, \dot{f}$ and $A$ of the unresolved sources (in black) and the resolved ones (in red), for $T_{\mathrm{obs}}=2$ years.}
    \label{fig:corner-all-res-f-fdot-A}
\end{figure*}

Figure \ref{fig:corner-all-res-f-fdot-A} shows the parameter space distributions of frequency $f$, frequency derivative $\dot{f}$ and amplitude $A$ for the unresolved (black) and resolved (red) sources, for $T_\mathrm{obs} = 2$ years. Resolved sources occupy higher frequencies ($\sim 1-5$ mHz) and higher amplitudes (above $\sim 10^{-23}$) compared to the unresolved population, reflecting the chirp-mass dependence of Eq. \ref{eq:amplitude-chirp-mass}: at fixed frequency, systems with larger $\mathcal{M}_c$ are easier to resolve. In the Figure, two distinct subpopulations are also clearly visible: one at intermediate frequencies and lower amplitudes, the other at lower frequencies and intermediate amplitudes. These correspond to two chirp mass subpopulations and will contribute differently to the background (we will discuss them in more detail in Section \ref{sec:confusion-background}).

Although no source in the catalog has exactly $\dot{f}=0$, many systems evolve so slowly that their frequency drift is effectively unmeasurable over the LISA mission lifetime: for $|\dot{f}| \leq 10^{-18}$ Hz/s, the total frequency evolution over four years is approximately $\sim 10^{-10}$ Hz, just below the frequency resolution $\Delta f = 1/T_\mathrm{obs} \sim 10^{-9}$. Throughout this work, such systems are classified as having $\dot{f} \simeq 0$. 

The resolved population presents a strong bias with respect to frequency evolution (Table \ref{tab:mchirp-fdot-res-all}). Monochromatic systems ($\dot{f} \simeq 0$) constitute 99.6\% of the catalog, but are almost never resolved: the make up only 0.5\%, 0.8\% and 1.6\% of the resolved catalog, for $T_\mathrm{obs}= 1,2,4$ years respectively. The resolved sample is instead drawn almost entirely from the rarer inspiraling ($\dot{f}>0$, 0.3\% of the catalog) and outspiraling ($\dot{f}<0$, 0.08\% of the catalog) populations. Their relative share within the resolved set shifts with observation time: the outspiraling fraction grows from 15.3\% to 20.1\% between 1 and 4 years, at the expense of the inspiraling fraction (84.2\% to 78.3\%), consistent with outspiraling binaries' systematically lower chirp mass placing more of them near the detection threshold, where they benefit most from the SNR gain of longer $T_\mathrm{obs}$.

\begin{table*}[htbp]
    \centering
    \caption{Number of sources in the catalog and resolved divided according to their frequency derivatives. Percentages in the Catalog and Resolved columns indicate the fraction of each type of sources in the catalog and in the resolved set respectively.}
    \label{tab:mchirp-fdot-res-all}
    \begin{tabular}{|c|cccc|}
        \hline
         & Catalog (\% total) & Resolved 1 yr (\% res) & Resolved 2 yrs (\% res) & Resolved 4 yrs (\% res) \\ \hline
        $\dot{f} <0$  & 12282 (0.08\%) &  686 (15.3\%) &  1322 (18.0\%) &  2371 (20.1\%)  \\
        $\dot{f} \simeq 0$ & 15169087 (99.6\%) & 23 (0.5\%) & 60 (0.8\%)   &  194 (1.6\%) \\
        $\dot{f} >0$ & 46231 (0.30\%) & 3772 (84.2\%)  &  5966 (81.2\%)    & 9246 (78.3\%)  \\ 
        \hline
        Total &  15227600   &   4481    &   7348    &   11811 \\
        \hline
    \end{tabular}
\end{table*}

\begin{figure}[]
    \centering
    \includegraphics[width=\linewidth]{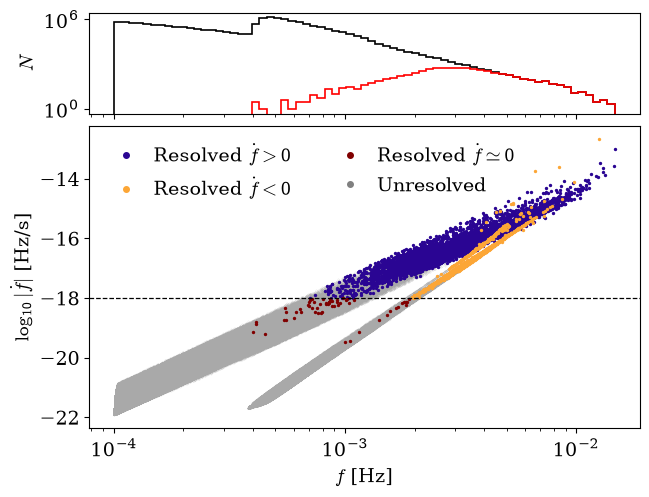}
    \caption{\textit{Top panel:} frequency distribution of all binaries in the catalog (black) and of the resolved (red) for $T_\mathrm{obs} = 2$ years. \textit{Bottom panel:} frequency derivative as function of frequency for unresolved sources (grey), and resolved ones with positive $\dot{f}$ (blue), negative $\dot{f}$ (orange) and $\dot{f}\simeq 0$ (maroon). The horizontal dashed line denotes the threshold of $|\dot{f}|$ below which a binary has $\dot{f}\simeq 0$.}
    \label{fig:f-fdot-all-resolved}
\end{figure}

The origin of this bias is visible in the $f-\dot{f}$ plane (bottom panel of Figure \ref{fig:f-fdot-all-resolved}). Inspiraling systems follow the GW-driven relation $\dot{f} \sim  f^{11/3}$ \cite{farmer_gravitational_2003}, and their comparatively large chirp masses and high frequencies make them the easiest population to detect. Monochromatic systems instead sit at low frequency, where Galactic confusion is stronger and their modest chirp mass keeps their SNR below threshold, despite their numerical dominance.
Outspiraling binaries occupy an intermediate position: mass-transfer episodes place them near LISA's most sensitive frequencies \cite{toubianaInteractingDoubleWhite2024}, but their typically small chirp masses (due to mass loss during interactions) give them lower amplitude than inspiraling systems at the same frequency \cite{marsh_mass_2004, kremer_accreting_2017, nelemans_short-period_2004}. This leaves them over-represented among resolved binaries relative to their catalog fraction, while remaining less detectable with respect to the inspiraling population. This catalog implements the same mass-transfer and tidal modeling as Toubiana et al. \cite{toubianaInteractingDoubleWhite2024}, just a different realization of the same underlying simulation. The excess of these sources between approximately 0.5 and 1.5 mHz in the top panel of Figure~\ref{fig:f-fdot-all-resolved} is therefore an expected consequence of sharing the same model, consistent with their reported accumulation around $\sim 1$ mHz.

\subsection{Confusion background} \label{sec:confusion-background}
In this section we characterize the confusion noise, and which sources dominate it.

\begin{figure}
    \centering
    \includegraphics[width=0.95\linewidth]{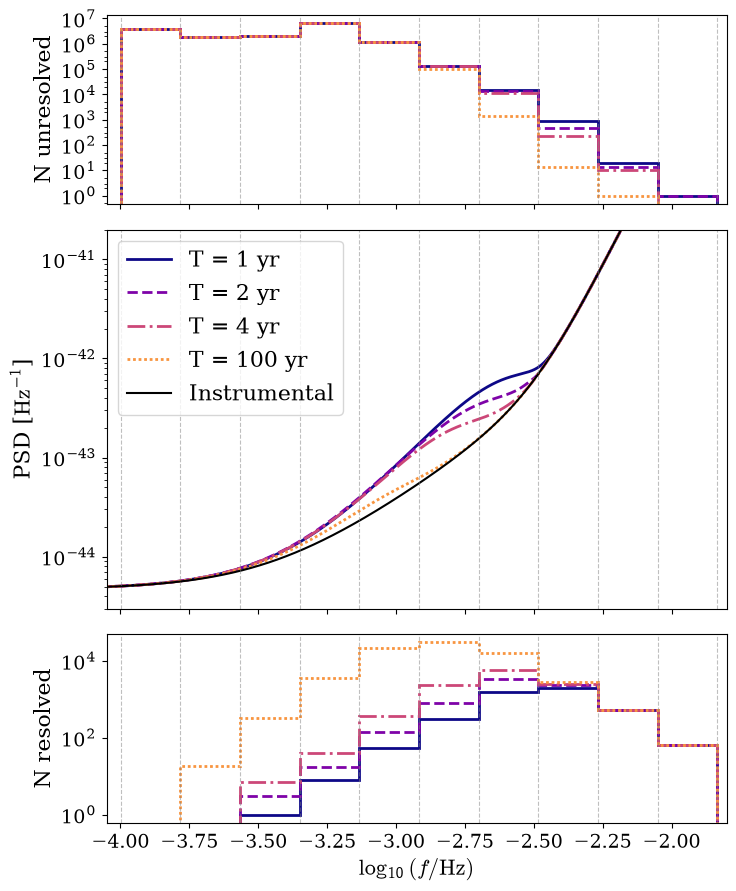}
    \caption{\textit{Top panel:} number of sources contributing to the confusion background per frequency band, for different $T_\mathrm{obs}$. \textit{Middle panel:} confusion PSD for 1 year (blue, solid curve), 2 years (purple, dashed curve), 4 years (pink, dash-dotted curve) and 100 years (yellow, dotted curve). The black curve is the instrumental noise. \textit{Bottom panel:} number of resolved sources per frequency band, for each $T_\mathrm{obs}$.}
    \label{fig:psds}
\end{figure}

Figure \ref{fig:psds} shows, for different $T_\mathrm{obs}$, the confusion PSD (middle panel) together with the number of sources contributing to the background and the number of resolved sources in each frequency band(top and bottom panels respectively). As expected, the confusion PSD decreases with longer $T_{\mathrm{obs}}$. 

In principle, given a sufficiently long observation time, all sources would eventually become resolvable. In practice, however, the mission lifetime is finite, and this is never achieved. To check this, an unrealistically long $T_\mathrm{obs}=100$ years was additionally tested: even in this limit, a residual confusion remains, indicating that there is an irreducible background due to sources being so numerous that they cannot be disentangled from each other. Moreover, it should be noted that, in this analysis, frequency resolution is assumed to scale as $\Delta f = 1/T_\mathrm{obs}$, and resolvability is determined solely by SNR, without accounting for confusion between individually resolved sources. In practice, this resolution cannot increase indefinitely, since data analysis is performed on shorter segments ($T\ll T_\mathrm{obs}$) that are later combined \cite{littenbergGlobalAnalysisGravitational2020}; moreover, sources with high SNR but very similar frequencies and amplitudes may still be difficult to disentangle. Neither effect is included here, and both remain areas for future work. As a consequence, the yellow dotted curve in Figure \ref{fig:psds} should be considered a lower limit: in realistic conditions, the residual background is expected to be higher, and effectively irreducible.

The number of resolved sources increases with $T_\mathrm{obs}$, while the number of unresolved ones decreases, and both span several orders of magnitude across the frequency range (top and bottom panels of Figure \ref{fig:psds}). Unresolved sources per bandgo from $\sim10^5$ at the lowest frequencies down to $\sim10$--$10^2$ at the highest; resolved sources go the other way, from essentially none at the lowest frequencies up to $\sim10^3$--$10^4$ at the highest. At lower frequencies, the fraction of unresolved sources remains almost unchanged by the increasing observation time: essentially the whole population there stays unresolved regardless of $T_\mathrm{obs}$. At intermediate frequencies, longer $T_\mathrm{obs}$ resolves considerably more sources, and this is where the four cases separate most clearly, most notably for $T_\mathrm{obs}=100$ years, where the resolved count reaches $\sim2\times10^4$ around $f\sim1$--$2$ mHz before converging back with the shorter-$T_\mathrm{obs}$ curves at the highest frequencies, where few sources remain in the catalog to resolve regardless of observation time.

\begin{figure}
    \centering
    \includegraphics[width=0.95\linewidth]{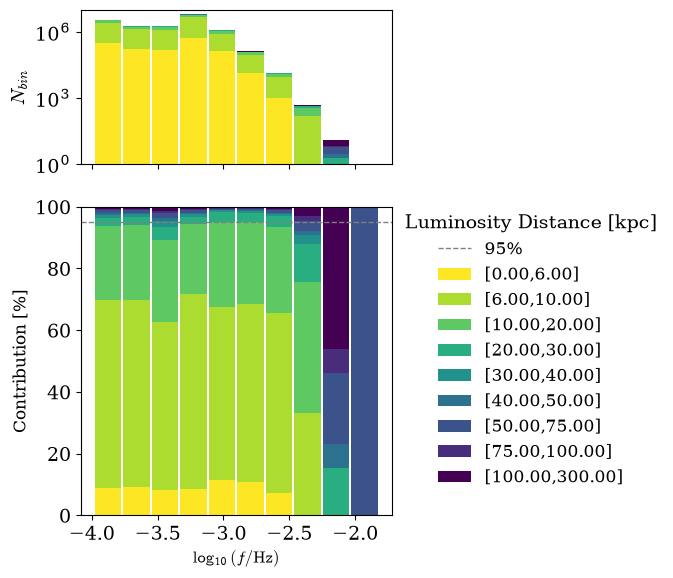}
    \caption{Distance contribution to the background per frequency band. The dashed line denotes the 95\% of the background power.}
    \label{fig:dist-contrib}
\end{figure}

Figure \ref{fig:dist-contrib} shows the distance contribution to the background per frequency band. Unresolved sources, which extend to up to 300 kpc from the Sun, are divided in distance shells, and their power contribution is added up to reach the 95\% of the total noise (dashed grey line). At frequencies below $\sim 3$ mHz, about 95\% of the background contribution originates from sources located within 20-30 kpc from the Sun. This is, in large part, simply a reflection of where the catalog itself is concentrated, rather than a distinct selection effect: a shell of 20 kpc around the Sun already encloses the entire dense bulge and most of the disk, and accounts for 93.6\% of the full catalog, so the background is dominated by nearby sources mostly because that is where most of the sources already are. In fact, the biggest contribution in most frequency bands comes from sources between 6 and 10 kpc, a distance shell including the bulge. We return to this point in Section \ref{sec:spatial-bias} when discussing resolved systems. At higher frequencies, nearby sources are increasingly likely to be resolved, meaning that the remaining background is dominated by distant systems. However, as shown in Figure \ref{fig:psds}, the confusion PSD itself becomes negligible above $\sim 3$ mHz, where the instrumental noise becomes the dominant component. This distance-based decomposition asks where the sources dominating the confusion budget are located. The following paragraphs turn to what they are, in terms of mass and amplitude rather than location. 

\begin{figure*}[t]
    \centering
    \includegraphics[width=0.8\linewidth]{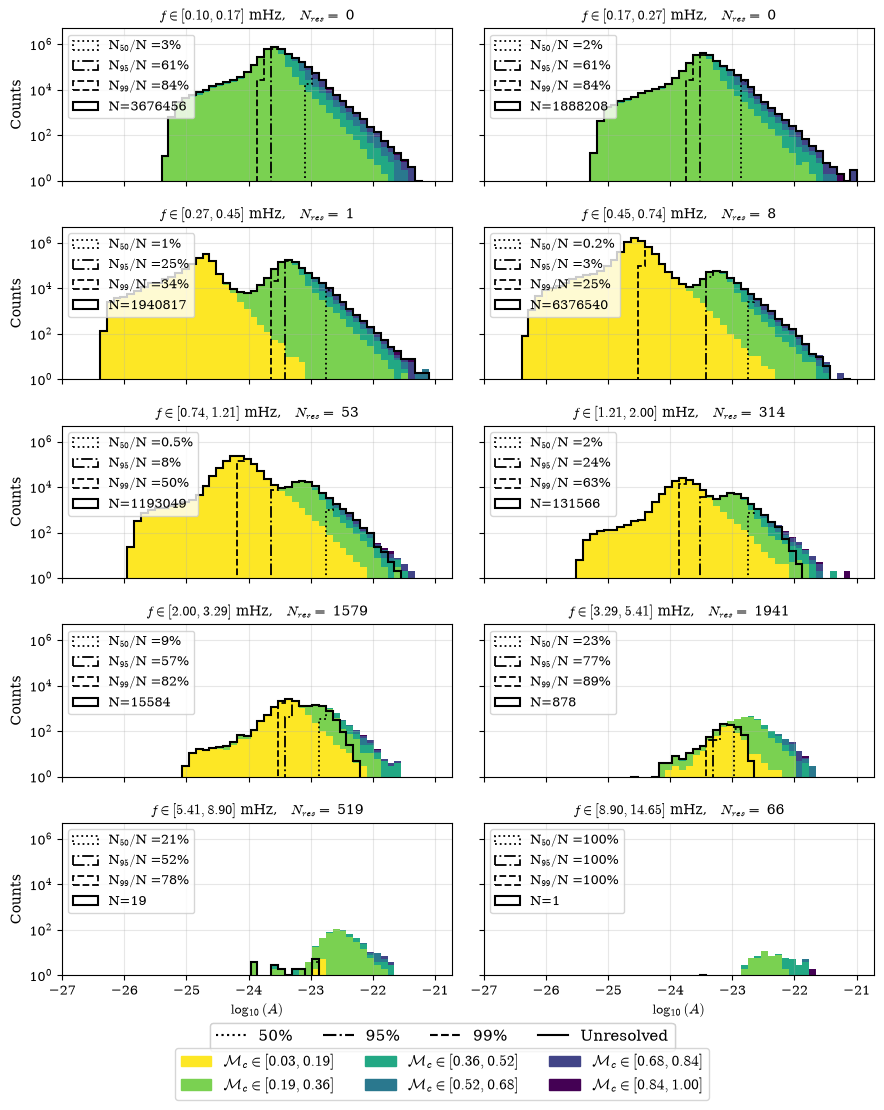}
    \caption{Amplitude distribution in each frequency band, sources shown as stacked histograms colored by chirp mass (in solar masses). The black outline encloses the full unresolved population, dashed and dotted lines mark the amplitude below which sources are excluded from the 50\%/95\%/99\% most powerful contributors to the confusion background. Resolved sources are included in the stacked distribution, outside the black outline, and their count, $N_\mathrm{res}$, is reported in each panel title. The fraction of unresolved sources contributing to each threshold and the total number of unresolved sources are given in each panel legend.}
    \label{fig:ampl-Mc-stacked}
\end{figure*}

The same chirp mass that drives detectability, $A \propto \mathcal{M}_c^{5/3} f^{2/3}/d_L$ (Section \ref{sec:detectability-bias}), also structures the population that remains unresolved: in a given frequency band, the amplitude distribution of unresolved sources is frequently bimodal (see Figure \ref{fig:ampl-Mc-stacked}), with two amplitude peaks corresponding directly to the two distinct chirp-mass sub-populations visible in Figure \ref{fig:corner-all-res-f-fdot-A}: a weaker one below $\mathcal{M}_c \sim 0.2 \ M_\odot$, on the lower end of the typical DWD range, and a louder one extending to higher masses (up to 1 $M_\odot$). This structure is not a compositional artifact: splitting the unresolved population by subtype and repeating the same per-band amplitude histograms shows the same two-peaks pattern within each subtype individually, simply occupying its own characteristic mass range.

To quantify how much of the confusion power each sub-population accounts for, we apply the $N_{99}$ diagnostic of Section \ref{sec:minimum-source-count-N99-A99} in each frequency band, and show the result in Figure \ref{fig:ampl-Mc-stacked}. 
The value of $N_{99}/N_\mathrm{unresolved}$ varies strongly with frequency, and its shape tracks the bimodality directly. In the lowest frequency band ($f\in[0.10,0.17]$ mHz), where the distribution is closest to unimodal, 99\% of the confusion power requires 84\% of the unresolved population. In the three bands with the highest source density, corresponding to those where the bimodality of the population is more pronounced, $N_{99}/N_\mathrm{unresolved} =$ 34\% ($f\in[0.27,0.45]$ mHz), 25\% ($f\in[0.45,0.74]$ mHz), and 50\% ($f\in[0.74,1.21]$ mHz). This means that a numerical minority accounts for the large majority of the background across this range. The minimum sits at $f \in [0.45, 0.74] $ mHz, in the densest band where the two peaks are more pronounced. Outside this range the bimodality is weaker, and at the higher bands this structure vanishes, and $N_{99}/N_\mathrm{unresolved}$ increases again.

The resolved and unresolved populations do not correspond to the same distribution. As shown in figure \ref{fig:ampl-Mc-stacked}, resolved sources occupy the highest amplitudes and chirp masses, continuous with the upper end of the louder sub-population. Only a small fraction of the low-chirp-mass population is resolved, leaving that subset heavily under-represented despite it constituting a large share of the unresolved sources in most bands. This has direct consequences for population inference: priors cannot be transferred from resolvable set to the whole DWD population.

\subsection{Spatial bias} \label{sec:spatial-bias}

In this section, we investigate how the $1/d_L$ scaling of the GW amplitude shapes the spatial distribution of the resolved population, and its impact on the angular power spectrum of resolved sources.

\begin{figure*}
    \centering
    \includegraphics[width=0.7\linewidth]{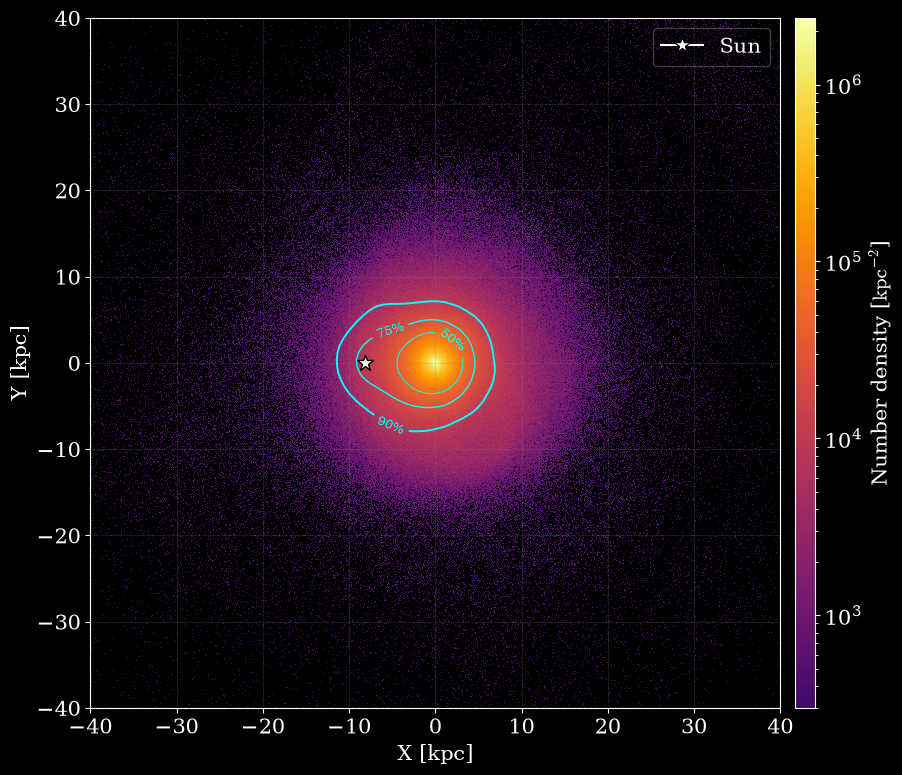}
    \caption{Face-on view of the catalog with contours enclosing 50\%, 75\% and 90\% of the resolved sources. The star denotes the position of the Sun.}
    \label{fig:density-XY}
\end{figure*}

The $\sim 1/d_L$ dependence of the amplitude (Eq. \ref{eq:amplitude-chirp-mass}) imposes a distance selection that makes sources in the proximity of the Sun more likely to be detected (see Appendix for details). However, despite this strong selection effect, the resolved population's overall shape still traces the bulge-dominated structure of the parent catalog, rather than being dominated by the nearby local disk. Figure \ref{fig:density-XY} shows the face-on view of the catalog with contours enclosing 50\%, 75\% and 90\% of the resolved sources: they are centered on the Galactic Bulge rather than on the Sun, but are visibly elongated toward the solar position, the 90\% contour extending to $X \simeq -12$ kpc on the Sun's side against $\simeq +8$ kpc on the far side. 

The bulge dominates through density, not proximity: farther from the Sun than the bulk of the local disk, and $\sim 20\times$ less likely to be detected than a source at 1 kpc, it still contains most resolved systems. The distance selection remains a smaller, secondary effect: it produces the elongation of the contours towards the Sun, and an over-representation within $\sim 3$ kpc. 95\% of resolved sources lie within 14 kpc from the Sun, a distance including only $\sim 86\%$ of the catalog, meaning that the resolved population is more tightly concentrated around the Sun than the full catalog.

\begin{figure*}
    \centering
    \includegraphics[width=\linewidth]{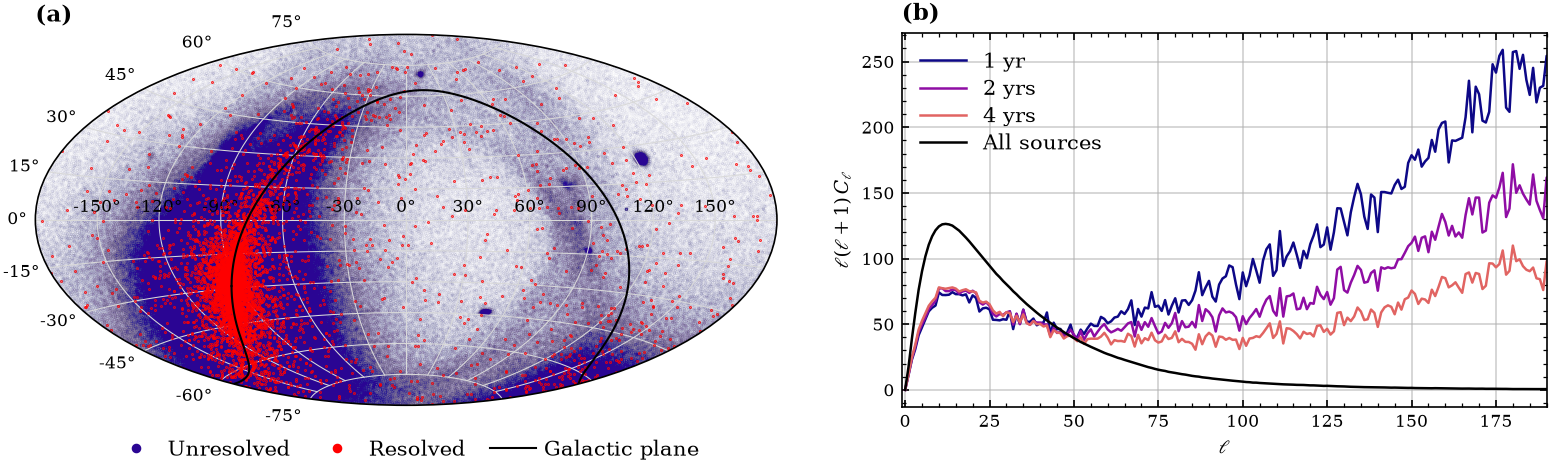}
    \caption{\textit{Panel a:} Sky distribution of resolved sources (red, $T_\mathrm{obs} = 2$ yrs) and unresolved sources (blue) in ecliptic coordinates. The black line denotes the equator. \textit{Panel b:} Angular power spectrum of all sources (black) and resolved sources for 1 year (blue), 2 years (purple) and 4 years (orange).}
    \label{fig:skyplot-powerspectrum}
\end{figure*}

At the opposite extreme from the bulge sits the Galactic halo: far more distant and far more diffuse than either the disk or the bulge, it offers a test of how complete the resolved catalog really is. We define the halo using the same criteria as Lamberts et al. \cite{lambertsPredictingLISAWhite2019} (Section \ref{sec:methods}). Above $\sim 4$ mHz, the catalog is nearly completely resolved even for halo sources, matching Lamberts et al.'s conclusion that LISA will produce a complete catalog down to a given frequency, halo included \cite{lambertsPredictingLISAWhite2019} (see Appendix for details). However, this depends strongly on the simulated catalog: like ours, Lamberts et al.'s simulation omits massive satellite galaxies such as the Magellanic Clouds, which are expected to host detectable systems \cite{korol_detectability_2018, korol_populations_2020}. The extragalactic satellites visible in Figure \ref{fig:skyplot-powerspectrum} are way less massive than real ones, but still provide an idea of what LISA will be able to resolve. 

The sky view of the simulated catalog is shown in the left panel of Figure~\ref{fig:skyplot-powerspectrum}, comparing resolved (red) and unresolved (blue) sources. Both populations peak toward the Galactic bulge, and some overdensities of unresolved sources are visible, corresponding to satellite galaxies. To compare the level of anisotropy between the resolved sources and the full catalog, we computed the angular power spectrum of both of these. This is shown in the right panel of Figure \ref{fig:skyplot-powerspectrum}, for all sources (black) and resolved ones for observation times of 1 (blue), 2 (orange)  and 4 years (pink). The angular power spectrum, where multipoles $\ell$ correspond to angular scale $\theta = 180^\circ / \ell$, reveals the typical angular scale of the galaxy. The peak of the full-catalog spectrum sits at $\ell = 12$, corresponding to $\theta= 15^\circ$, matching the observed angular extent of the Galactic bulge of the Milky Way \cite{simionParametricDescription3D2017}. This is a useful sanity check of the simulated catalog.

The spectra of the resolved sources confirm a stronger concentration of sources towards the bulge than the full catalog, even though their peak is less pronounced: this is a consequence of shot noise,  which arises from the finite number of resolved sources \cite{jenkinsEstimatingAngularPower2019}, and adds power roughly uniformly across multipoles and therefore dilutes the peak's prominence. The rise at high multipoles $\ell$ is also due to the shot noise; longer observation times reduce this effect at high multipoles as more sources are resolved, and therefore the shot noise in the resolved catalog is reduced. 

Together, these results show that, while the population's broad structure still traces the bulge, like the full catalog, its detailed spatial distribution is not a fair representation of the underlying binary population. LISA will preferentially resolve systems in the bulge, where source density is higher, more than what the underlying population's bulge concentration would suggest. It will also sporadically resolve systems outside the bulge and off the Galactic plane. The angular power spectrum comparsion in Figure~\ref{fig:skyplot-powerspectrum}  highlights a potential spatial bias between resolved sources and galactic binary noise which should be taken into account when informing search and inference algorithms of each of these components. For example, the spherical harmonic decomposition of the galactic binary noise component has already been targeted in proposed map-making strategies for LISA~\cite{banagiri_mapping_2021}.

\subsection{Composition bias} \label{sec:composition-bias}

In this section we examine how the detectability bias identified in Section \ref{sec:detectability-bias} translates into an over- or under- representation of specific WD compositional subtypes in the resolved catalog. Following Lamberts et al. \cite{lambertsPredictingLISAWhite2019}, we perform an analogous analysis of the compositional contribution of the resolved population. The catalog used in this analysis is a different realization of the same underlying simulation (Section \ref{sec:methods}), generated with different initial conditions: this makes the two studies' astrophysical model directly comparable, but the specific realization, and therefore exact source counts will differ. We therefore compare trends rather than precise numbers. We also note that Lamberts et al.'s catalog underwent an additional source removal step not applied to ours: differences we find between the two studies' compositional trends should be interpreted with this in mind.

Table \ref{tab:fraction-types-resolved-catalog} shows the fraction of each subtype in the full catalog and in the resolved population for $T_{\mathrm{obs}} = 1,2,4$ years. Every subtype's share of the resolved set differs from its share of the full catalog: He-CO is under-represented, decreasing from 54.8\% of the catalog to $\sim48\%$ of the resolved set, while He-He, CO-CO, and Ne-X are all over-represented. Ne-X shows this most strongly, roughly doubling its share of the resolved set relative to its catalog fraction. 

As $T_{\mathrm{obs}}$ increases, He-He systems' share in the resolved population increases from 24.8\% (1 yr) to 26.2\% (4 yrs). This same behaviour, with a larger increase, was also found by Lamberts et al. \cite{lambertsPredictingLISAWhite2019}, since these systems lie close to the detectability threshold and benefit the most from the increase in SNR with observation time.
\begin{table}[htbp]
    \centering
    \caption{Fraction of each type of binary in the catalog and in the resolved set, for different $T_{\mathrm{obs}}$.}
    \label{tab:fraction-types-resolved-catalog}
    \begin{tabular}{|c|c|c c c|}
        \hline
        Type & \% Catalog  & \multicolumn{3}{c|}{\% Resolved }  \\ 
             &         & 1yr & 2yr & 4yr   \\ \hline
        He-CO & 54.8\% & 48.6\% & 48.2\% & 48.6\%  \\
        He-He & 22.7\% & 24.8\% & 25.5\% & 26.2\%  \\
        CO-CO & 21.0\% & 23.1\% & 23.0\% & 22.2\%  \\
        Ne-X  & 1.5\%  & 3.4\%  & 3.2\%  & 3.0\%   \\ \hline
    \end{tabular}
\end{table}

\begin{figure*}[htbp]
    \centering
    \includegraphics[width=\linewidth]{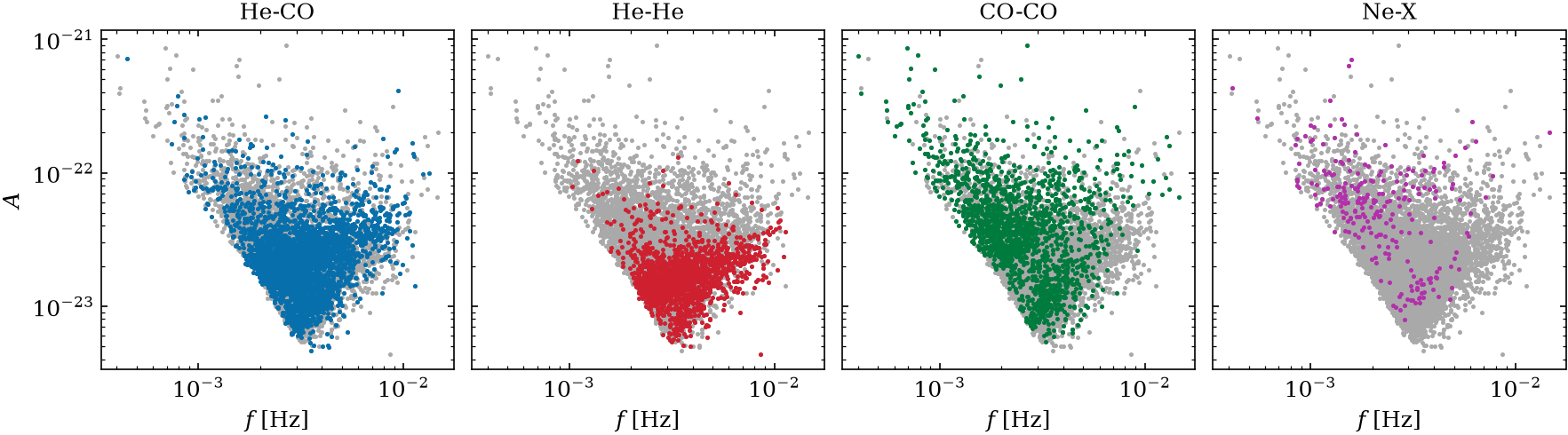}
    \caption{Amplitude-frequency distribution for resolved sources of each binary type. In grey are all resolved sources ($T_\mathrm{obs} = 2$ yrs), while colored points highlight the different types: He-CO (blue), He-He (red), CO-CO (green) and Ne-X (purple).}
    \label{fig:resolved-comp-f-A-distrib}
\end{figure*}

This pattern is explained by each subtype's typical frequencies and chirp masses, which set its typical amplitude and therefore its detectability. This is visible directly in the frequency-amplitude distribution of resolved sources (Figure \ref{fig:resolved-comp-f-A-distrib}): CO-CO and Ne-X binaries, with the largest average chirp masses, occupy the highest-amplitude region and are correspondingly the most over-represented subtypes. He-CO spans a broad range of both frequency and amplitude, which keeps its resolved fraction ($\sim$ 48\%) closer to its catalog share (54.8\%) than any other subtype's. He-He binaries, despite their smaller chirp mass, are over-represented in the resolved share, due to their typically higher frequencies.

This compositional bias compounds the parameter-space and spatial biases identified in the previous sections: the resolved catalog will over-represent nearby, bulge-concentrated, high-chirp-mass systems, and this must be accounted for when using LISA observations to infer properties of the Milky Way's binary population.

\subsection{Time evolution of the confusion} \label{sec:time-evolution}
In this section, we investigate the yearly modulation of the confusion background.

\begin{figure*}[ht]
    \centering
    \includegraphics[width=0.9\linewidth]{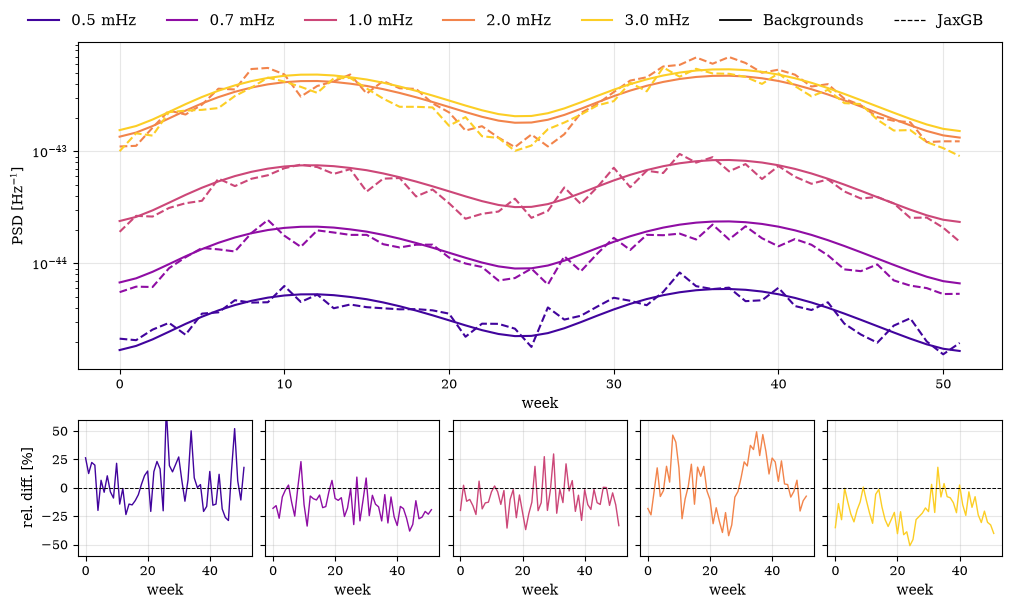}
    \caption{\textit{Top panel:} yearly modulation of the confusion PSD at different frequencies. Solid curves are computed using the response from $\texttt{backgrounds}$, dashed curves are the weekly computed PSDs using data produced with \texttt{JaxGBaccurate}. \textit{Bottom panel:} relative difference between the weekly \texttt{backgrounds} modulation of the PSD and the computed one, for each frequency.}
    \label{fig:psd-modulation-bkg-jaxgb}
\end{figure*}

The top panel of Figure \ref{fig:psd-modulation-bkg-jaxgb} shows the level of the confusion PSD as a function of time at five representative frequencies over one year, computed with \verb|backgrounds| (solid lines) and using \verb|JaxGBaccurate| for the time-domain data generation (dashed lines). The modulation of both methods shows the two expected peaks over the year, with one higher than the other. In the bottom panel, the relative difference between the two methods is plotted. At higher frequencies, the low number of sources contributing to the background introduces shot noise, visible as irregular spikes. Peaks and wells in the computed modulation are accentuated by the concentration of these sources toward the Galactic center, especially at high frequencies. At lower frequencies this effect is averaged out by the larger number of binaries in each frequency bin. 

The magnitude of the PSD increases with frequency, before decreasing again above $\sim3$ mHz, tracing the shape of the confusion noise shown in figure \ref{fig:psds}. 

The cyclostationarity of the confusion noise needs to be taken into account when analyzing or simulating LISA data, as it introduces a non-negligible amplitude modulation that affects source subtraction and inference. Moreover, accounting for a time-dependent confusion improves parameter estimation of resolved sources \cite{buscicchioFirstYearLISA2025}, and could also be a method of disentangling the Galactic foreground from the instrumental noise and other isotropic noise sources, such as extragalactic backgrounds \cite{boileauGravitationalWaveBackgroundExtragalactic2025, perego_assessing_2026, buscicchio_test_2025}. A simplified approach like the one shown in this work can be used as a baseline for the expected behaviour, with more complete approaches, such as wavelet decomposition, left to future work \cite{cornish_time-frequency_2020, digman_lisa_2022}. 

\section{Discussion} \label{sec:discussion}

\subsection{Origin of the bias}

The results presented in Section \ref{sec:results} consistently point to the same conclusion: the population of DWD binaries resolved by LISA is a biased subset of the underlying population. 

The selection effects of the resolved population are all directly related to the evolutionary history of the binaries: the driving quantities for detection are the frequency, the chirp mass and the frequency derivative of the systems. The first two parameters affect directly the GW amplitude and therefore the SNR of the binary, leading to different selection effects across the parameter space. 

The dominance of inspiraling systems in the resolved population reflects the fact that these binaries are preferentially drawn from the higher chirp-mass sub-population identified in Section \ref{sec:confusion-background} (see also the Appendix), whose larger GW amplitude drives them above the detection threshold. Binaries undergoing and surviving mass-transfer episodes (outspiraling, $\dot{f}<0$), by contrast, have lower chirp masses and correspondingly smaller amplitudes: they remain over-represented relative to their catalog fraction, since mass-transfer episodes place them near LISA's most sensitive frequencies, but far less compared to inspiraling systems, which stay the easiest population to detect. Monochromatic systems, which constitute the majority of the catalog, are naturally characterized by lower frequencies, placing them below the detection threshold and making them the least likely of all three groups to be resolved.

The same chirp-mass dependence structures the confusion background itself (Section \ref{sec:confusion-background}). The unresolved population is not homogeneous: in frequency bands where the two chirp-mass subpopulations are clearly separated, a numerical minority of higher-mass systems accounts for the great majority of the confusion, while the far more numerous, largely outspiraling low-mass population contributes comparatively little power (See the Appendix).

The spatial bias reflects two distinct effects (Section \ref{sec:spatial-bias}). The $1/d_L$ scaling of the GW amplitude means closer systems are more easily resolved, and since the Sun sits within the Galactic disk, a large share of the catalog is already nearby. Independently, the Galactic bulge is intrinsically the densest region of the Galaxy in source number, and this density advantage outweighs the modest extra distance relative to the surrounding disk, even though the Sun itself is not part of the bulge. Both effects concentrate the resolved population toward the inner Galaxy, a concentration also visible in the angular power spectrum.

The same mechanism drives the compositional bias (Section \ref{sec:composition-bias}). CO-CO and Ne-X binaries, which have the largest typical chirp masses among the four types, are over-represented in the resolved population relative to their catalog fraction. Ne-X systems show the most extreme effect, appearing at roughly twice their catalog fraction ($\sim 3\%$ resolved against $\sim 1.5\%$ in the catalog). He-CO binaries, which dominate the catalog at 54.8\%, are under-represented in the resolved set: unlike the other three subtypes, which all present a larger share of inspiraling systems, He-CO is split evenly between inspiraling and outspiraling binaries, which limits how much of its population reaches the higher chirp masses that favor detection. He-He binaries, despite typically smaller chirp masses, are also over-represented in the resolved set relative to the catalog, because a larger share of them are inspiraling rather than mass-transferring. The resolved fraction of He-He binaries also increases progressively with $T_\mathrm{obs}$.

Taken together, these biases form a coherent picture. LISA's resolved DWDs will be dominated by massive, inspiraling systems that are nearby and concentrated in the Galactic bulge. The stable, low-frequency systems that make up the majority of the Milky Way's DWD population will remain almost entirely in the confusion foreground.

\subsection{Limitations}

The main limitation of this analysis is the idealized subtraction of the resolvable sources during the iterative pipeline. The procedure assumes perfect knowledge of source waveforms, perfect noise characterization, and the ability to subtract resolved sources without residuals or source confusion. In practice, waveform mismodeling, overlapping signals at the same frequency, and non-stationary noise or glitches will all degrade the subtraction. The number of resolved sources reported here should therefore be interpreted as optimistic upper bounds.

A further limitation is how much we can actually infer about the underlying population from the confusion background alone. The $N_x/A_x$ diagnostic (Section \ref{sec:minimum-source-count-N99-A99}) ranks sources by amplitude only, and different combinations of sources can produce the same total confusion power in a given frequency band. This degeneracy means that individual source properties, or the detailed shape of the underlying population, cannot be recovered directly from the background alone. However, this does not necessarily rule out population-level inference: hierarchical Bayesian approaches, which constrain population statistics rather than individual sources, may still work despite this degeneracy. This is an active area of ongoing work: \cite{toubiana_framework_2026} develop a hierarchical framework that connects the resolved and unresolved populations consistently to the underlying astrophysical population, \cite{criswell_foundation_2026} address a closely related question specifically within the LISA Global Fit, and \cite{santiInferringPopulationProperties2026} present a prototype for relating the measured foreground directly to the astrophysical population. A complementary approach bypasses the resolved/unresolved split altogether through simulation-based inference applied directly to the data \cite{srinivasan_simulation-based_2025}.

Finally, the results depend on the simulated catalog. While the catalog used here represents a realistic simulation of the Milky Way's DWDs, the binary-evolution model behind it is probably inaccurate in its details: current population-synthesis codes are not yet constrained well enough to make precise predictions. The catalog also shares the same underlying simulation and astrophysical assumptions as Lamberts et al. \cite{lambertsPredictingLISAWhite2019} (Section \ref{sec:methods}), just with different initial conditions, so the agreement we find with their trends (Section \ref{sec:composition-bias}) reflects consistency within the same model, not an independent check of the assumptions behind it. We expect the qualitative conclusions of this work, the existence of the detectability biases we identify, to hold regardless of these uncertainties; the specific percentages reported throughout will change as the underlying binary-evolution models improve.

\subsection{Implications for LISA}

These results have direct consequences for how LISA data should be interpreted. Any inference about the underlying population from the resolved set must account for the detection biases characterized in this work: treating the resolved catalog as a representative sample would systematically overestimate the fraction of massive, inspiraling systems concentrated in the Galactic bulge, at the expense of the far more numerous, low-mass, stable population that dominates the Galaxy.

The angular power spectrum of the resolved sources (Section \ref{sec:spatial-bias}) provides an additional probe of this spatial concentration. The peak at $\ell = 12$, corresponding to an angular scale of $15^\circ$, is consistent with the observed angular extent of the Milky Way bulge \cite{simionParametricDescription3D2017}, and reflects the strong bulge contribution to the resolved population established above.

Moreover, the cyclostationarity of the foreground introduces a time-dependent component that, in principle, can help distinguish it from other, more stationary noise sources \cite{pozzoli_cyclostationary_2025}. This is relevant for global-fit and inference pipelines (Section \ref{sec:time-evolution}): a stationary treatment of the noise risks biasing parameter estimation for sources near the detection threshold, in the frequency bins and epochs where the yearly modulation is largest. Explicitly taking into account this modulation may improve both the completeness and the parameter accuracy of such pipelines.

Taken together, these results matter beyond this specific catalog. An accurate model of what LISA does and does not resolve, as well as the confusion background left behind, is a prerequisite for realistic global-fit pipelines and source subtraction, and it sets the benchmark against which future simulated Galactic DWD catalogs should be judged. Ultimately, the resolved catalog will be LISA's primary observational window onto the Galactic DWD population: the astrophysics extracted from it, from formation channels to the structure of the Milky Way, is only as reliable as our understanding of how that catalog was shaped in the first place.

\section{Acknowledgements}

We thank Astrid Lamberts and Alexandre Toubiana for providing the simulated catalog underlying this work and for their detailed feedback on this work. We thank Michele Vallisneri for supervising the Master's thesis this work is built on, and Maria Radisch for her collaboration and support throughout the project. We thank Quentin Baghi for a helpful discussion about \texttt{backgrounds} and Maude Le Jeune for a helpful discussion at an early stage of this work. 
This work was supported by the Swiss LISA Data Processing Center (DPC@CH) at ETH Zurich.

\section*{Appendix}

\subsection*{Decomposing the confusion background}

\begin{figure}[h]
    \centering
    \includegraphics[width=\linewidth]{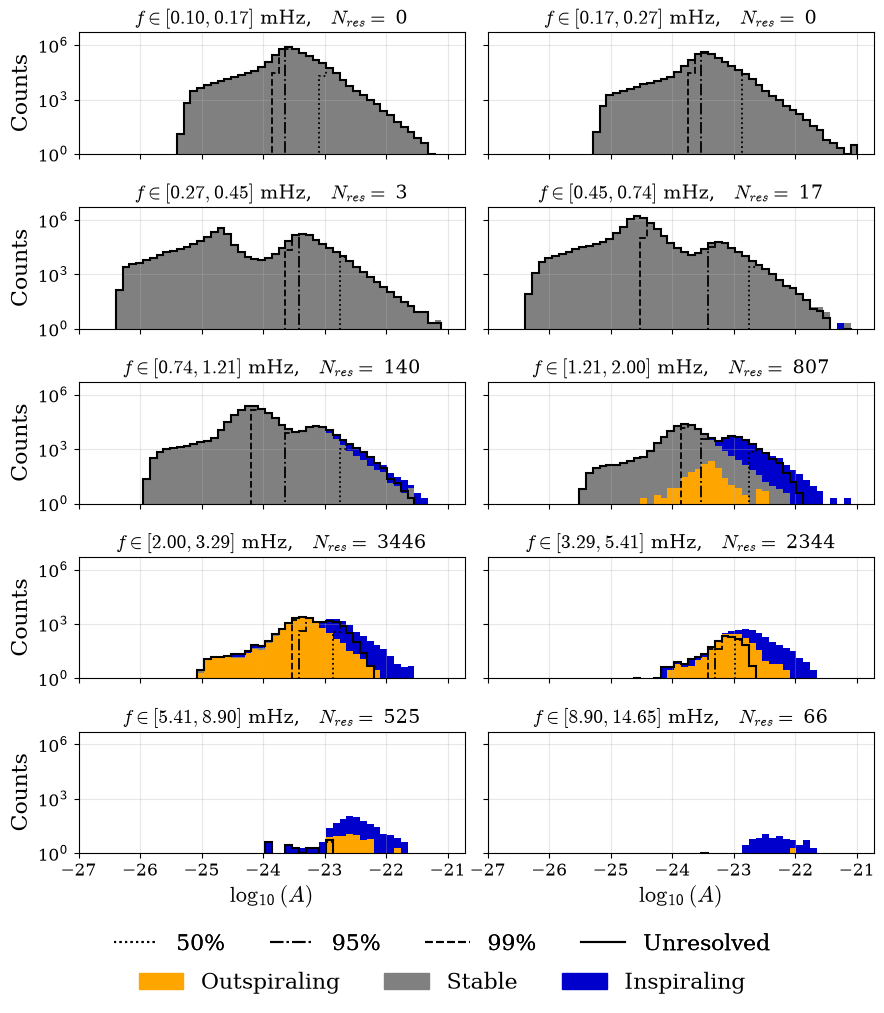}
    \caption{Amplitude distribution in each frequency band, sources shown as stacked histograms colored by frequency derivative. The black outline outlines the full unresolved population, dashed and dotted lines mark the amplitude below which sources are excluded from the 50\%/95\%/99\% most powerful contributors to the confusion background. Resolved sources are included in the stacked distribution, outside the black outline, and their count, $N_\mathrm{res}$, is reported in each panel title.}
    \label{fig:appendix-ampl-fdot-stacked}
\end{figure}

\begin{figure}[h]
    \centering
    \includegraphics[width=\linewidth]{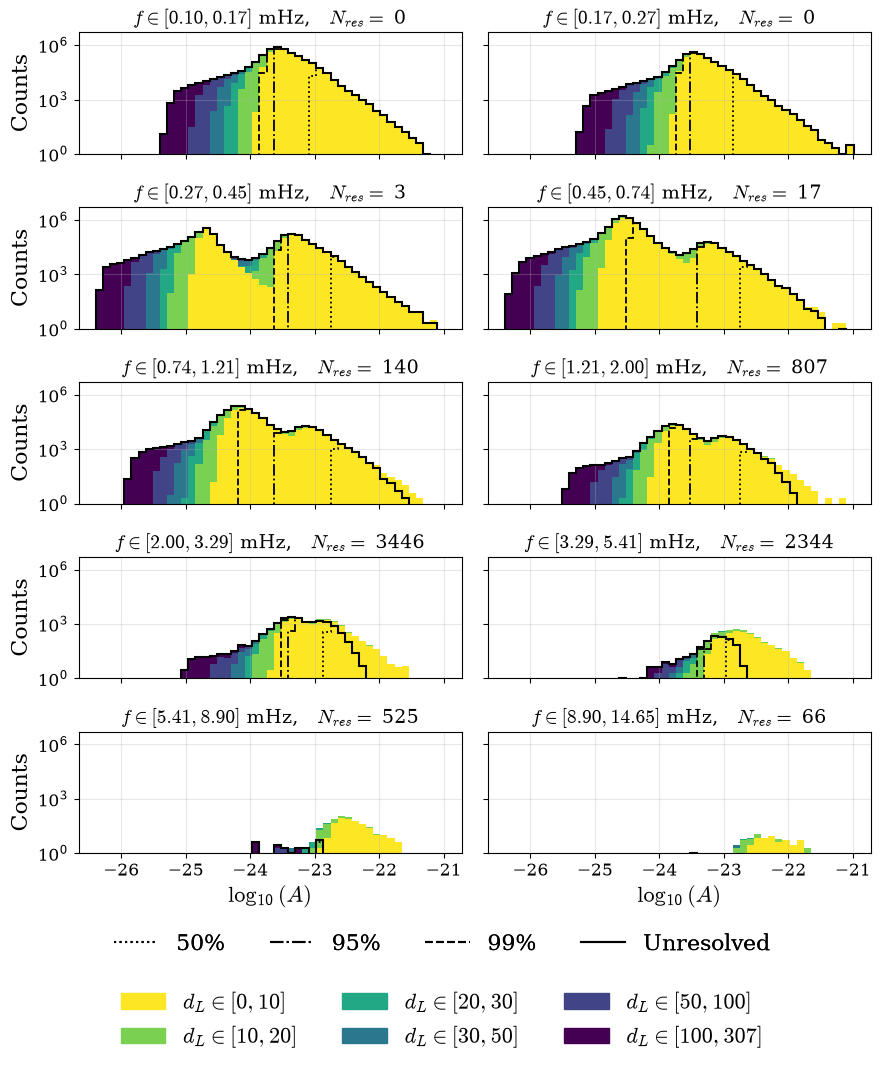}
    \caption{Amplitude distribution in each frequency band, sources shown as stacked histograms colored by luminosity distance (in kpc). The black outline outlines the full unresolved population, dashed and dotted lines mark the amplitude below which sources are excluded from the 50\%/95\%/99\% most powerful contributors to the confusion background. Resolved sources are included in the stacked distribution, outside the black outline, and their count, $N_\mathrm{res}$, is reported in each panel title.}
    \label{fig:appendix-ampl-dl-stacked}
\end{figure}

To quantify how much of the confusion background each population accounts for, we adopt the ordering strategy described in Section \ref{sec:minimum-source-count-N99-A99}. The results are shown in Figures \ref{fig:ampl-Mc-stacked}, \ref{fig:appendix-ampl-fdot-stacked} and \ref{fig:appendix-ampl-dl-stacked}, with the amplitude distribution colored by chirp mass, frequency derivative and luminosity distance, respectively.

Coloring the amplitude distribution by $\dot f$'s sign (Figure \ref{fig:appendix-ampl-fdot-stacked}) shows that the lower-$\mathcal{M}_c$, lower-amplitude peak is preferentially populated by $\dot f \le 0$ systems, while the higher-$\mathcal{M}_c$ peak that dominates the background is preferentially $\dot f>0$. 

Figure \ref{fig:appendix-ampl-dl-stacked} shows the amplitude distribution colored by $d_L$: most of the sources are located in the bulge of the galaxy, between 0 and 10 kpc from the Sun (colored in yellow), and therefore contribute the most to the background power. The two populations are not characterized by different distance distributions, meaning that $d_L$ doesn't affect the background contribution other than the usual scaling $A \propto 1/d_L$.

\subsection*{Distance selection function}

\begin{figure}
    \centering
    \includegraphics[width=\linewidth]{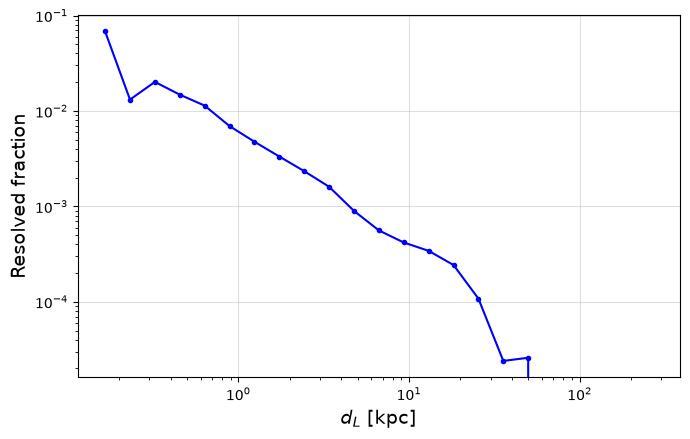}
    \caption{Fraction of sources in the catalog that are resolved at a given luminosity distance from the Sun.}
    \label{fig:dl-scaling-fraction}
\end{figure}

In section \ref{sec:spatial-bias} we discussed how the $\sim1/d_L$ scaling affects the resolved sources. Figure \ref{fig:dl-scaling-fraction} shows the fraction of sources that are resolved as a function of distance from the Sun. This is a strong selection effect, with the fraction of resolved sources dropping by a factor of $\sim 20$ between 1 and 10 kpc from the Sun. Despite this, the Galactic Bulge remains dominant due to its incredibly high source density, as explained in Section \ref{sec:spatial-bias}.

\subsection*{Halo sources in the resolved catalog}

\begin{figure}[htbp]
    \centering
    \includegraphics[width=\linewidth]{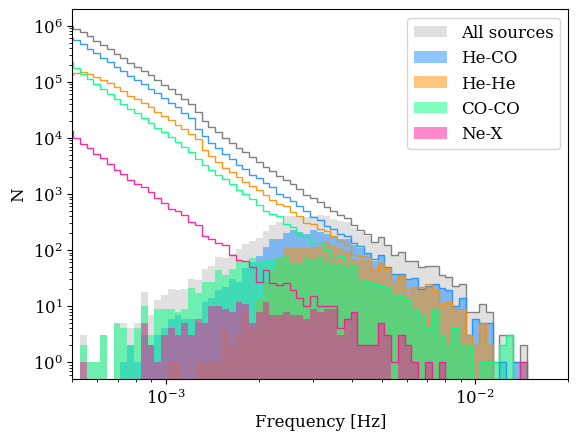}
    \caption{Frequency distribution of the resolved sources for $T_\mathrm{obs} = 2$ yrs. In grey are all resolved sources, while colored histograms correspond to the different binary types. The contours indicate the frequency distribution of all the sources in the catalog.}
    \label{fig:freqs-resolved-types}
\end{figure}

Figure \ref{fig:freqs-resolved-types} shows the frequency distribution of all sources and of resolved ones for each binary subtype. Above $\sim 4$ mHz, almost all sources are resolved, including halo sources. Only 41 halo sources are resolved for $T_\mathrm{obs}=2$ yrs, corresponding to 0.56\% of the resolved set, compared to the 5\% in the full catalog, since halo sources typically have lower frequencies. The halo's composition also differs from the full catalog, with a higher share of He-He binaries, explaining the typically lower frequencies and amplitudes of the halo.

\bibliography{references_fixed}

\end{document}